\documentclass[12pt,a4paper]{article}
\usepackage{amsmath}
\usepackage{amssymb}
\usepackage{amsthm}
\usepackage{float}
\usepackage{amsfonts}
\usepackage{graphicx}
\usepackage{verbatim}
\usepackage[left=2cm,right=2cm,top=3cm,bottom=2.5cm]{geometry}
\usepackage[numbers]{natbib}
\usepackage[utf8]{inputenc}
\usepackage[usenames,dvipsnames,svgnames]{xcolor}
\usepackage[colorlinks=true,
      linkcolor=red,
      urlcolor=gray,
      citecolor=blue]{hyperref}

\def\myalign#1{%
  \def\trule{\noalign{\smallskip\hrule\medskip}}
  \def\nebc{\nearrow\bigcup}
  \def\sebc{\searrow\bigcup}
  \def\pminf{{}_{-\infty}|^{+\infty}}
  \let\Inf\infty
  \def\amp{&} 
  \vbox{\mathsurround0pt\openup1\jot
    \halign{%
      &$\displaystyle##\hfil\tabskip0pt$&\amp##\tabskip1em\crcr
      \noalign{\hrule height1pt\smallskip}#1\noalign{\smallskip\hrule height1pt}\crcr}}}
      
\begin{document}

\begin{center}
 \textbf{Inflation constraints for classes of $f(R)$ models}
\end{center}
\hfill\newline
\begin{center}
Joseph Ntahompagaze$^{1,2}$, Jean Damasc\`{e}ne Mbarubucyeye$^{1}$, Shambel Sahlu$^{2,3}$ and Amare Abebe$^{4}$\\
Email: ntahompagazej@gmail.com\\
\end{center}
\hfill\newline
$^{1}$ Department of Physics, College of Science and Technology, University of Rwanda, Rwanda\\
$^{2}$ Astronomy and Astrophysics Division, Entoto Observatory and Research center, Ethiopian Space Science and Technology Institute\\
$^{3}$Department of Physics, College of Natural and Computational Science, Wolkite University, Ethiopia\\
$^{4}$ Center for Space Research, North-West University, South Africa \\


\noindent In this paper, we explore the equivalence between two theories, namely $f(R)$ and scalar-tensor theories of gravity. We use this equivalence to explore several $f(R)$ toy models focusing on the inflation epoch of the early universe.
The study is done based on the definition of the scalar field in terms of the first derivative of $f(R)$ model. We have applied the slow-roll approximations during inflationary parameters consideration. The comparison of the numerically computed inflationary parameters with the observations is done. We have inspected that some of the $f(R)$ models produce numerical values of $n_{s}$ that are in the same range as the  suggested values from observations. But for the case of the tensor-to-scalar ratio $r$, we realized that some of the considered $f(R)$ models suffer to produce a value which is in agreement with the observed values for different considered space parameter.\\
\newline
$keywords:$ $f(R)$ gravity; scalar-tensor; inflation; scalar field; cosmology.

\section{Introduction}
The wide exploration of how modified gravity theories respond to modern cosmological observations is getting much attention nowadays.  This exploration is part of the ongoing validation and viability tests being carried out against some of the proposed modified theories of gravity \cite{scalar1,capozziello2006cosmological}.
The motivation to propose the modified gravity theories is also based on what Einstein's theory of relativity failed to explain in the modern cosmological observations such as cosmic acceleration observed two decades ago \cite{1}. The standard model of cosmology is based on the Einstein's theory of relativity. The modified gravity theories get attention because of the failure of General Relativity (GR) to explain the cosmic acceleration and other outstanding problems in the early universe \cite{debono2016general}. The most accepted hypothesis is that dark energy, a fluid which does not interact with photons, with a negative pressure, is responsible for this acceleration of the expansion of the universe \cite{frieman2008dark}. But several works suggest that using the modification of GR, one can produce the cosmic history that one could get with dark energy \cite{trodden2007cosmic}. On the other hand, the recent detection of gravitational waves \cite{abbott2016observation} puts GR to a level that brings complications to the modifiers of this theory.  

\noindent Some of the modified gravity theories currently having a wide attention are $f(R)$ gravity and scalar-tensor theories of gravity.
One can modify the action so that instead of having the Ricci scalar $R$ (for GR case), one has its generalization $f(R)$. Consequently, the resulting field equations are modified too. This type of modified gravity theory was proposed firstly by Buchdahl in 1970. Later, it was explored in several works (see for example in \cite{barrow1983stability} how the stability analysis of different $f(R)$ models is treated). 
Different ways of exploring the $f(R)$ theory of gravity is increasingly getting attention because of different motivations. In \cite{alexeyev2017astronomical}, the study about how $f(R)$ theories, especially the Starobinsky model can be compared to GR on astrophysical scales (galaxies scales) was initiated. But one can also study how $f(R)$ theories are equivalent to scalar-tensor theories of gravity \cite{whitt1984fourth,jakubiec1988theories}. 
For instance in \cite{ntahompagaze2017f}, the authors considered different $f(R)$ models in the context of scalar-tensor theories of gravity. In that work,
the scalar field considered was based on its definition used in \cite{scalar5}, that is $\phi=f'-1$, in such a way that it vanishes for $f(R)=R$, that is in GR. 
There are other different ways of defining the scalar field when one is dealing with the $f(R)$ theory, see for example \cite{amin2016viable,chakraborty2016solving,clifton2005power}.

\noindent In the current paper, we consider a different way of defining the scalar field based on the work done in \cite{7}. For $f(R)=R$, the scalar field defined is a constant. The relationship between the scalar field and the derivative of $f(R)$ is given as:
\begin{equation}\label{scalarfield}
\phi=f'.
\end{equation}
\noindent We focus on the early universe, mostly the inflation epoch. During this epoch, cosmological expansion was accelerated exponentially \cite{guth1993inflation}. A couple of approaches to treat the behavior of the inflation are put in place. For example in \cite{liddle1994formalizing}, the proposition of applying the slow-roll approach was made. The application of the slow-roll approximation is so wide that one can compute some of the inflationary parameters \cite{easther2006implications,slow3,liddle1994formalizing}. The proposition to apply constant-roll is also proposed \cite{motohashi2017constant}. In this paper we are interested in applying the slow-roll approximations and try to numerically compute some of the inflationary parameters (say scalar spectral index $n_{s}$ and tensor-to-scalar ratio $r$) with the purpose to compare them with the results of the recent cosmological survey known as the Planck Survey. 

\noindent It is getting common to compute inflationary parameters in modified gravity theories. For example in \cite{sebastiani2014nearly}, after exploring how the Einstein's frame can be used to reconstruct $f(R)$ models by specification of the potential, many $f(R)$ models have been analyzed and in return respective values of $n_{s}$ and $r$ are computed for a given number of e-fold $N$. 
In \cite{myrzakulov2015inflationary}, the computation of both $n_{s}$ and $r$ is made where the consideration of higher-derivative quantum gravity that contains Gauss-Bonnet terms is covered. They obtained values of $n_{s}$ and $r$ that are compatible with the observations (Planck results).
The consideration of both models that contain logarithimic terms and others with exponential terms is done in \cite{odintsov2017unification}, where they obtained values of $n_{s}$ and $r$ that are in the range predicted by the observations, they used constant-roll approximations. In that work, it has been concluded that $f(R)$ model that have exponential terms, the one studied there, can produce values of $n_{s}$ and $r$ that are in good range if the constant-roll approximation \cite{motohashi2017constant} is preferred than slow-roll approximation.

\noindent Concerning Jordan and Einstein frames, to move from one frame to another frame, one has to perform the so-called conformal transformation. Say if one wants to move from Jordan frame to Einstein frame, the metric $g_{\mu\nu}$ has to be transformed and consequently the Ricci scalar $R$ is also transformed \cite{magnano1994physical,faraoni1999einstein}. One can see how this works in \cite{sebastiani2014nearly}. In \cite{postma2014equivalence} it is shown how the two frames describe the same physics. One can also see the dynamical equivalence between the two frames covered in \cite{saltas2011dynamical}. However, in \cite{capozziello2006cosmological} it has been shown that for some models the mathematical equivalence between the two frames does not imply the physical equivalence. However, for the inflationary epoch this equivalence of the two frames is affected by the appearence of singularities or sometimes from the shift from decceleration to acceleration \cite{bahamonde2016correspondence,bahamonde2017deceleration}. For instance in \cite{bahamonde2017deceleration}, it has been shown that when one is dealing with $f(R)$ gravity, the situation that allows the acceleration phase in one frame will not necessarily imply the same phase in the other frame. However this problem is cured by making sure that the conformal invariant quantities are used rather than non-conformal invariants \cite{bahamonde2017deceleration}.
In the current paper, we will use the Jordan frame and because of that, the specification of the potential will not be necessary a priori, it will rather be derived from a definition provided in \cite{scalar5}.

\noindent It is however useful here to mention some of the possible ends of inflation epoch. One of the possibilities is to make the potential steep such that it puts an end to the slow-roll era \cite{liddle2003long}. The other one is to involve a second field as an hybrid to the standard scalar field so that the new field brings inflation to an end \cite{linde1994hybrid,linde1997hybrid,lyth1996more}. The study done in \cite{elizalde2008reconstructing} shows how one can reconcile the inflationary epoch with the late-time accelerated epoch. This was done in single field and mutilple fields scenarios.

\noindent The adopted spacetime signature is $(-,+,+,+)$ and the Ricci scalar $R$ is defined as:
\begin{equation}
R=R_{\mu\nu}g^{\mu\nu}\; , \label{Ricciscalar1}
\end{equation}
where $R_{\mu\nu}$ is the Ricci tensor. This tensor is defined as
\begin{equation}
R_{\mu\nu}=R^{\alpha}_{\mu\alpha\nu}\; , 
\end{equation}
where the Riemann tensor $R^{\alpha}_{\beta\mu\nu}$ is given as
\begin{equation}
R^{\alpha}_{\beta\mu\nu}=\Gamma^{\alpha}_{\beta\mu,\nu}-\Gamma^{\alpha}_{\beta\nu,\mu}
+\Gamma^{\alpha}_{\tau\nu}\Gamma^{\tau}_{\beta\mu}-\Gamma^{\alpha}_{\tau\mu}\Gamma^{\tau}_{\beta\nu}\; .
\label{Riemann1}
\end{equation}
In the above equation, the symbol $\Gamma^{\mu}_{\nu\lambda}$ is called the Christoffel symbol defined in terms of the metric $g_{\mu\nu}$ as
\begin{equation}
\Gamma^{\mu}_{\nu\lambda}=\frac{1}{2}g^{\mu\alpha}(g_{\alpha\nu,\lambda}+g_{\alpha\lambda,\nu}
-g_{\nu\lambda,\alpha})\; . \label{Christoffel1}
\end{equation}
This paper is organized as follows. In the following section, we provide the basic equations to be used in the subsequent sections. In Section \ref{DIFFERENTFRS}, we consider and study different $f(R)$ models, Section \ref{DISCUSSIONSSS} is devoted for discussions, and the last Section for conclusions.

\section{Basics equations}
The action for $f(R)$ gravity theories is given as:
\begin{equation}\label{fRaction}
 \mathcal{A}= \frac{1}{2\kappa} \int d^{4}x\sqrt{-g}\left[ f(R)+2\mathcal{L}_{m}\right]\; ,
\end{equation}
where $\kappa= 8\pi G$ and $\mathcal{L}_{m}$ is the matter Lagrangian. 
Scalar-tensor theories of gravity currently have gained more attention in the context of exploring the extra-degree of freedom of modified gravity \cite{ scalar1,scalar2,7}. The Brans-Dicke theory of gravity is one of the scalar-tensor models with a constant coupling parameter $\omega$. Its action is written as
\cite{scalar1,7}
\begin{equation}\label{DICKE}
I_{BD}=\int d^{4}x \sqrt{-g}\big[\phi R-\frac{\omega}{\phi}\nabla_{\mu}\phi\nabla^{\mu}\phi+\mathcal{L}_{m}(\Psi,g_{\mu\nu})\big].
\end{equation} 
The exploration of the equivalence between $f(R)$ and ST theory dates back to 1988, where Barrow and Cotsakis explored the quadratic $f(R)$ as a subclass of Brans-Dicke ST theory \cite{barrow1988inflation}. Later, much works in relation to this equivalence were done \cite{scalar1,scalar5,7,scalar3}. Here, we briefly conceptualize this equivalence as follows. For a given $f(R)$ Lagrangian, one can define an auxiliary field $\chi$ such that it is a function of the scalar field $\phi$ as $\chi(\phi)$. Then, the potential $V(\phi)$ can be written as \cite{7}
\begin{equation}\label{RedefPOTENTIAL}
V(\phi)=\chi(\phi)\phi-f(\chi(\phi))\;.
\end{equation}
Then the action presented in Eq. \eqref{fRaction} becomes
\begin{equation}
S_{f(R)}=\frac{1}{2\kappa}\int dx^{4}\sqrt{-g}\Big[f(\chi)+f'(\chi)(R-\chi)+\mathcal{L}_{m}\Big]\;.
\end{equation}
We therefore use Eqs. \eqref{RedefPOTENTIAL} and \eqref{scalarfield} to have the action presented in Eq. \eqref{fRaction} rewritten as
\begin{equation}
S_{f(\phi)}=\frac{1}{2\kappa}\int dx^{4}\sqrt{-g}\Big[\phi R-V(\phi)+\mathcal{L}_{m}\Big]\;.
\end{equation} 
This equation is equivalent to Eq. \eqref{DICKE} for a vanishing coupling parameter $\omega$. The manifestation of the $f(R)$ is usually hidden in the definition of the potential $V(\phi)$.
In $f(R)$ gravity theory, one can vary the action with respect to the metric only (results in metric formalism), or with respect to the other parameters (this results in Palatini formalism and metric-affine formalism depending on the variation type). Here  we will be interested in the metric formalism.
The action in Eq. \eqref{fRaction} produces the field equation presented in \cite{7,amare4} as
\begin{equation}
G_{\mu\nu}=\frac{1}{f'}\big[T^{m}_{\mu\nu}+\frac{1}{2}g_{\mu\nu}(f-Rf')+\nabla_{\nu}\nabla_{\mu}f'-g_{\mu\nu}\nabla_{\sigma}\nabla^{\sigma}f'\big],\label{eqfr0}
\end{equation}
where $f=f(R)$, $f'=\frac{df}{dR}$ and $T^{m}_{\mu\nu}=-\frac{2}{\sqrt{-g}}\frac{\delta (\sqrt{-g}\mathcal{L}_{m})}{\delta g^{\mu\nu}}$ is the energy-momentum tensor (EMT) of standard matter. 
The field equations from the action in Eq. \eqref{fRaction} are given in Ref. \cite{7} as:
\begin{equation}
 G_{ab}=\frac{\kappa}{\phi}T^{m}_{ab}+\frac{1}{\phi}\left[\frac{1}{2} g_{ab}\left(f-\phi R\right)
 +\nabla_{a}\nabla_{b}\phi-g_{ab}\square \phi \right],\label{frstt2a}
\end{equation}
where $\square =\nabla_{c}\nabla^{c}$ is the covariant d'Alembert operator. The EMT for the scalar field is given as
\begin{equation}
T^{\phi}_{ab}=\frac{1}{\phi}\left[\frac{1}{2} g_{ab}\big(f-\phi R\big)+\nabla_{a}\nabla_{b}\phi-g_{ab}\square \phi\right].
\end{equation}
The scalar field $\phi$ obeys the Klein-Gordon equation \cite{scalar5}
\begin{equation}
\square \phi -\frac{1}{3}\big(2f-\phi R+T^{m}\big)=0, \label{KG}
\end{equation}
where $T^{m}$ is the trace of the matter EMT. The effective potential term is given as \cite{scalar5}
\begin{equation}
V'(\phi)=\frac{dV}{d\phi}=\frac{1}{3}\big(2f-\phi R\big)\;.\label{pot}
\end{equation} 
Note that the definition of $\chi(\phi)$ is done in such a way that the integration of Eq. \eqref{pot} produces the same result as Eq. \eqref{RedefPOTENTIAL} for a vanishing constant of integration.
We focus our attention on the early universe, the inflationary epoch. In this epoch, the universe was dominated by scalar field. Within this context, one can consider the situation where the scalar field was slowly evolving over the potential. The slow-roll assumptions is based on such an assumption. 
There are two conditions leading to slow-rolling \cite{liddle1994formalizing}. The first one says that the square of the time derivative of the slow-rolling scalar
field has to be smaller than the slow-rolling scalar field potential. This is mathematically:
\begin{equation}
\big(\frac{d\phi}{dt}\big)^{2}<V(\phi). \label{slow1}
\end{equation}
The second condition is about the second-order time derivative which is conditioned to be smaller than the derivative of the potential with 
respect to the scalar field $\phi$. This is
\begin{equation}
2\big|\frac{d^{2}\phi}{dt^{2}}\big|<|V'(\phi)|.\label{slow2}
\end{equation}
The parameters defined below can be obtained from \cite{liddle2000cosmological,lyth2009primordial}:
\begin{equation}\label{EPSILONN}
\epsilon \approx \frac{1}{2\kappa^{2}}\left(\frac{V'}{V}\right)^{2},
\end{equation}
\begin{equation}\label{ETAA}
\eta \approx \frac{1}{\kappa^{2}}\left(\frac{V''}{V}\right)\;.
\end{equation}
The scalar spectral index $n_{s}$ and tensor-to-scalar ratio $r$ are given as
\begin{equation}\label{NSS}
n_{s}=1-6\epsilon+2\eta\;,
\end{equation}
\begin{equation}\label{RRR}
r=16\epsilon,
\end{equation}
respectively.
Once the relation between the scalar field $\phi$ and the Ricci scalar $R$ is established, we can obtain the inflation potential $V(\phi)$ from $V'(\phi)$ by performing the integration and the related expression such as $V''(\phi)$. If we have these expressions, we can get $\epsilon(\phi)$ and $\eta(\phi)$ so that in return compute $n_{s}$ and $r$. In the following we only present the results obtained after numerical computations for different $f(R)$ models.
\section{Some $f(R)$ models}\label{DIFFERENTFRS}
In this section we use the definition presented in Eq. \eqref{scalarfield} and the derivative of the potential presented in Eq. \eqref{pot} to obtain the expression for the potential $V(\phi)$. We use the result of the potential $V(\phi)$ and its first and second derivatives, $V'(\phi)$ and $V''(\phi)$ respectively, to obtain the inflationary parameters that are presented in Eqs. \eqref{EPSILONN}-\eqref{ETAA} to be able to obtain the final expressions of the spectral index and the tensor-to-scalar ratio presented in Eqs. \eqref{NSS} and \eqref{RRR}. We have considered different types of $f(R)$ models ranging from power law to exponential models. The motivations behind the consideration of these models are: (i) They are proven to be viable models that are compatible with cosmological observations. 
(ii) Most of them may pass solar system tests. (iii) They manifest a feature that they can mimic dark energy hypothesis when one is dealing with the past and current cosmic expansion of the universe. Besides that the current exercise aims to show that the observations can constrain the $f(R)$ gravity theories. 
\subsection{Model 1: $f(R)=\beta R^{n}$}\
This model is the simplest generalization of $GR$, is widely studied and has well-known cosmological (exact) solutions.
The model was first considered in  \cite{barrow1983stability} in the study of stability analysis of $f(R)$ models. For this model,
\begin{equation}
f'=\beta nR^{n-1}\;,
\end{equation}
and therefore
\begin{equation}
\phi=\beta nR^{n-1}\;.
\end{equation}
We get the Ricci scalar as function of the scalar field as
\begin{equation}
R(\phi)= \left( \frac{\phi}{\beta\,n} \right)^{\frac {1}{n-1}} \;.
\end{equation}
Then $f(\phi)$ is given as
\begin{equation}
f(\phi)=\beta  \left( \frac{\phi}{\beta\,n} \right)^{\frac {n}{n-1}}  \;.
\end{equation}
Then the equation for the derivative of the potential is given as
\begin{equation}
V'(\phi)=-\frac{\phi}{3}\left( \frac{\phi}{\beta\,n} \right)^{\frac {1}{n-1}}+\frac{2\beta}{3}\left( \frac{\phi}{\beta\,n} \right)^{\frac {n}{n-1}}\;,
\end{equation}
and integrating yields
\begin{equation}
V(\phi)=-\frac{\phi( n-1) }{3(2n-1)} \left( \phi\left( \frac{\phi}{\beta n} \right)^{\frac{1
}{n-1}}\right)-2\beta\left( \left( \frac{\phi}{\beta\,n} \right)^{\frac {n}{n-1}} \right) +D_{1}\; ,\label{PotentialRnmodel1}
\end{equation}
where $D_{1}$ is a constant of integration.
  \begin{table}[h!]
\centering
\caption{Model 1: $f(R)=\beta R^{n}$}
\begin{tabular}{|c|c|c|c|c|c|c|c|}  \hline
	Set&$D_{1}$&	$\beta$&	n&	$\phi$&	$n_{s}$&	$r$\\ \hline
	&0&	10& 	1.99&	0.75&& \\
	I&&&&to 1.0&	0.967$\pm$  0.006&0.236$\pm$ 0.033\\ \hline
	&200 to&&&&&	\\	
	II &1200~~&	0.1& 	1.2&	1.1&	0.96798$\pm$  0.015&0.313$\pm$ 0.065\\ \hline
	&10&	0.07 & 	1.3&	1.1&	0.9704$\pm$  0.005&0.214$\pm$ 0.018\\ 
   III &&to	0.105&&&&\\	 \hline
	&10&	0.5& 	1.3 &	1.1&	0.9683$\pm$  0.009&0.169$\pm$ 0.061 \\
	IV&&&to 1.85&&& \\\hline
	&10&	1& 	1.2 &	1.15&	0.9683$\pm$  0.005&0.035$\pm$ 0.007 \\
	V&&&to 1.24&&& \\\hline
	&10&	1& 	0.9 &	1.02&	0.9693$\pm$  0.0092&0.019$\pm$ 0.008 \\
	VI&&&&to 1.09&& \\\hline
	&2&	1& 	0.99 &	1.055&	0.9669$\pm$  0.0107&0.0003$\pm$ 0.0002 \\
	VII&&&&to 1.065&& \\\hline			
\end{tabular}\label{Model1}
\end{table}

\subsection{Model 2: $f(R)=\alpha R+\beta R^{n}$}\label{linearMODELLLL}
This model is one of the linear extension of GR with an additional power law term \cite{sebastiani2014nearly}. For vanishing $\beta$, GR is recovered with appropriate consideration of the running constant $\alpha$. However, some of the features of this model can be elaborated. For example, for $n=1$, the two terms are combined to produce GR-like model for $\beta\neq 0$. This model is widely studied and its consideration has been done in \cite{SanteCarloni2}, where dynamical system analysis of this model was investigated. The first derivative is
\begin{equation}
f'=\alpha+\beta nR^{n-1}\;,
\end{equation}
and hence
\begin{equation}
\phi=\alpha+\beta nR^{n-1}\;.
\end{equation}
We get the Ricci scalar as a function of the scalar field as
\begin{equation}
R(\phi)=\left( {\frac {\phi-\alpha}{\beta\,n}}
 \right)^{\frac {1}{n-1}}.
\end{equation}
Then $f(\phi)$ is given as
\begin{equation}
f(\phi)=\alpha\left( \frac {\phi-\alpha}{\beta\,n}
 \right)^{\frac {1}{n-1}}+\beta\left( \frac {\phi-\alpha}{\beta\,n}
 \right)^{\frac {n}{n-1}} \;.
\end{equation}
The potential  $V(\phi)$ is given as
\begin{equation}
\begin{split}
&V(\phi)=\frac {\beta\,n(2\alpha-\phi)}{ 3\left( n-1 \right) ^{-1}+1} {\left( {
\frac {\phi-\alpha}{\beta\,n}} \right) ^{ \left( n
-1 \right) ^{-1}+1}}
-\,{
\frac {2\alpha\,\beta\,n}{3(2\,n-1)}{\left( -\frac {\alpha-\phi}{\beta\,n}
 \right)^{\frac {n}{n-1}} }}\\
&+{\frac {{\beta}^{2}{n}^{2}}{3 \left(  \left( 
n-1 \right) ^{-1}+1 \right)  \left(  \left( n-1 \right) ^{-1}+2
 \right) } \left( {\frac {\phi-\alpha}{\beta\,n}} \right) ^{ \left( n-1 \right) ^{-1}+2}}\\
&+\,{\frac {2\alpha\,
\beta}{3(2\,n-1)}{\left( \frac {\phi-\alpha}{\beta\,n}
 \right)^{\frac {n}{n-1}} }}+\,{\frac {2\beta\, \left( n-1 \right) 
\phi}{3(2\,n-1)}{\left( \frac {\phi-\alpha}{\beta\,n}
 \right)^{\frac {n}{n-1}} }}+D_{2}\; ,
\end{split}
\end{equation}
where $D_{2}$ is a constant of integration.
  \begin{table}[h!]
\centering
\caption{Model 2: $f(R)=\alpha R+\beta R^{n}$}
\label{Model2}
\begin{tabular}{|l|c|c|c|c|c|c|c|c|} \hline
	Set&$D_{2}$&	$\alpha$&	$\beta$&	n&	$\phi$&	$n_{s}$&	$r$ \\ \hline 
	&0&	0.1& 	1.991&	1.4&	1.15 to&&\\
	I&&&&& 1.45&	0.969$\pm$0.005& 0.175$\pm$0.016 \\  \hline 
	&0&	0.008&&&&&\\
	II&& to 0.25& 	2&	1.5&	1.1&	0.965$\pm$0.002& 0.182$\pm$0.012 \\  \hline 
	&10&	0.01& 	0.1008&&&&\\
	III&&& to 0.104&	1.3&	1.1&	0.9736$\pm$0.0008& 0.204$\pm$0.003 \\  \hline 
	&10&	0.01& 	0.05&	1.3&&&\\
	IV&&&& to 1.4&	1.01&	0.9709$\pm$0.011& 0.204$\pm$0.055 \\  \hline 
	&1.4&&&&&& \\
	V&to 4.6&	0.01& 	0.05&	1.4&	1.01&	0.967$\pm$0.003& 0.202$\pm$0.013 \\	 \hline 	
\end{tabular}
\end{table}

\subsection{Model 3: $\alpha\,{e}^{\lambda\,R}$}

This model is treated in \cite{SanteCarloni2} where the stability analysis using dynamical system is performed. But in \cite{bamba2008future} different form of exponential models that are similar to this one are treated in that paper. With other models, the way this model responds to the finite-time future singularity as it was treated in \cite{bamba2008future}. One can notice that using taylor expansion, the polynomial $f(R)$ can be obtained. Because of that, its reduction to GR can be achieved. But its exponential form makes the analysis easier. However, the model-dependent constants $\alpha$ and $\lambda$ are non-zero. Because of that, we prefer to also study how it responds to the cosmological inflation. Here, we have the first derivative of the model given as
\begin{equation}
f'=\alpha\,\lambda\,{e}^{\lambda\,R}\;,
\end{equation}
so that
\begin{equation}
\phi=\alpha\,\lambda\,{e}^{\lambda\,R}\;.
\end{equation}
The Ricci scalar takes the form
\begin{equation}
R(\phi)=\ln  \left( {\frac {\phi}{\alpha\,\lambda}} \right) .
\end{equation}
Then $f(\phi)$ is given as
\begin{equation}
f(\phi)=\alpha\, \left( {\frac {\phi}{\alpha\,\lambda}} \right) ^{\lambda} \;.
\end{equation}
The potential  $V(\phi)$ is given as
\begin{equation}
V(\phi)=-\frac{1}{6}\,{\phi}^{2}\ln  \left( {\frac {\phi}{\alpha\,\lambda}} \right) +\frac{1}{
12}{\phi}^{2}+\,{\frac {2{\alpha}^{2}\lambda}{3(\lambda+1)} \left( {
\frac {\phi}{\alpha\,\lambda}} \right) ^{\lambda+1}}+D_{3}\;,
\end{equation}
where $D_{3}$ is a constant of integration.

  \begin{table}[h!]
\centering
\caption{Model 3: $f(R)=\rm{\alpha}e^{\lambda R}$}
\label{Model3}

\begin{tabular}{|c|c|c|c|c|c|c|c|} \hline
Set&$D_{3}$&$\alpha$ & $\rm{\lambda}$&$\phi$ &  $n_{s}$     & $r$    \\      \hline
&0&0.005 &     1.3 &&&\\
I&&&to 3.2      &1.01 &  0.969 $\pm$0.002  & 0.149 $\pm$ 0.001  \\  \hline	
&1 to&&&&&\\
 II&48&0.005 & 3.1&1.01 &  0.967 $\pm$0.003  & 0.187 $\pm$ 0.014  \\  \hline
&0.5&0.052 &&&&\\
III&&to 0.066 & 4.1&1.01 &  0.967 $\pm$0.004  & 0.236 $\pm$ 0.017  \\  \hline
&0.5&0.05 & 4.1&1.1 &&\\
IV&&&&to 1.5 &  0.965 $\pm$0.005  & 0.209 $\pm$ 0.036   \\  \hline
\end{tabular}
\end{table}
\subsection{Model 4: $R+\alpha\,\ln  \left( {\frac {R}{{\mu}^{2}}} \right) +\beta\,{R}^{m}$}
This model is an extension of GR with couple of terms. The two terms, namely logarithmic and power-law terms have a great contribution to how this model may produce different scenarios in cosmology. For example, if the series exapansion of logarithmic term is done, its combination with the power law term may reduce to GR-like model with appropriate consideration of the running constants and some truncations of higher order contributions.
The model was considered in \cite{nojiri2004modified,nojiri2011unified,nojiri2017modified}. However in \cite{nojiri2004modified}, the inflation analysis of this model was made using the definition of the scalar field addressed with the natural logarithimic approach. In \cite{nojiri2011unified}, the authors considered different conditions for which this model can be viable as a model that can describe the evolution of the universe, where they considered $m=2$. In this paper, we will keep $m=2$ for simplicity without loss of generality. The first derivative with respect to $R$ of this model is
\begin{equation}
f'=1+{\frac {\alpha}{R}}+2\,\beta\,R
\end{equation}
so that the scalar field becames
\begin{equation}
\phi=1+{\frac {\alpha}{R}}+2\,\beta\,R\; .
\end{equation}
Here, if one solves for $R(\phi)$, two different roots are obtained as
\begin{equation}
R(\phi)={\frac {\phi-1+\sqrt {-8\,\beta\,\alpha+{\phi}^{2}-2\,\phi+1}}{
4\beta}}\; ,\label{ROOT1}
\end{equation}
and 
\begin{equation}
R(\phi)=-\,{\frac {-\phi+1+\sqrt {-8\,\beta\,\alpha+{\phi}^{2}-2\,\phi+1}}{
4\beta}}\; .\label{ROOT2}
\end{equation}
We will name root1 from Eq. \eqref{ROOT1} and root2 from Eq. \eqref{ROOT2}.
From root1, we have $f(\phi)$ given as
\begin{equation}
\begin{split}
f(\phi)=&{\frac {\phi-1+\sqrt {-8\,\beta\,\alpha+{\phi}^{2}-2\,\phi+1}}{
4\beta}}+\alpha\,\ln  \left({\frac {\phi-1+\sqrt {-8\,\beta\,
\alpha+{\phi}^{2}-2\,\phi+1}}{4\beta\,{\mu}^{2}}} \right)\\
& +{
\frac { \left( \phi-1+\sqrt {-8\,\beta\,\alpha+{\phi}^{2}-2\,\phi+1}
 \right) ^{2}}{16\beta}}\; .
 \end{split}
\end{equation}
With the potential $V(\phi)$ being
\begin{equation}
\begin{split}
V(\phi)=&-{\frac {\phi}{12\beta}}+{\frac {{\phi}^{2}}{24\beta}}+\Big[-\frac{1}{24\beta}(\phi-1)-\frac{2\alpha}{3}\Big]\sqrt {-8\,\beta\,\alpha+{\phi}^{2}-2\,\phi+1} \\
&-\frac{\alpha\phi}{3}\big(1+4\ln  \left( 2 \right)\big)-\alpha\,\ln  \left( \phi-1+\sqrt {-8\,\beta\,\alpha+{\phi}^{2}-2\,\phi
+1} \right)\\ 
&+\frac{2}{3}\ln  \left( {\frac {\phi-1+\sqrt {-8\,\beta\,\alpha+{\phi}^{2}-2
\,\phi+1}}{\beta\,{\mu}^{2}}} \right) \phi\,\alpha
+D_{4}\; ,
 \end{split}
\end{equation}
where $D_{4}$ is a constant of integration.
From root2, we have $f(\phi)$ given as
\begin{equation}
\begin{split}
f(\phi)&={\frac {\phi-1-\sqrt {-8\,\beta\,\alpha+{\phi}^{2}-2\,\phi+1}}{4\beta}}+\alpha\,\ln  \left( {\frac {\phi-1-\sqrt {-8\,\beta\,
\alpha+{\phi}^{2}-2\,\phi+1}}{4\beta\,{\mu}^{2}}} \right)\\
& +{
\frac { \left( -\phi+1+\sqrt {-8\,\beta\,\alpha+{\phi}^{2}-2\,\phi+1}
 \right) ^{2}}{16\beta}}
\; .
\end{split}
\end{equation}
With the potential $V(\phi)$ being
\begin{equation}
\begin{split}
V(\phi)=&-{\frac {\phi}{12\beta}}+{\frac {{\phi}^{2}}{24\beta}}+\Big[-\frac{1}{24\beta}(\phi-1)+\frac{2\alpha}{3}\Big]\sqrt {-8\,\beta\,\alpha+{\phi}^{2}-2\,\phi+1} \\
&-\frac{\alpha\phi}{3}\big(1+4\ln  \left( 2 \right)\big)+\alpha\,\ln  \left( \phi-1+\sqrt {-8\,\beta\,\alpha+{\phi}^{2}-2\,\phi
+1} \right)\\ 
&+\frac{2}{3}\ln  \left( {\frac {\phi-1-\sqrt {-8\,\beta\,\alpha+{\phi}^{2}-2
\,\phi+1}}{\beta\,{\mu}^{2}}} \right) \phi\,\alpha
+D_{5}\; ,
 \end{split}
\end{equation}
where $D_{5}$ is a constant of integration.

  \begin{table}[h!]
\centering
\caption{Model 4: $R+\alpha\,\ln  \left( {\frac {R}{{\mu}^{2}}} \right) +\beta\,{R}^{2}$, root1}
\label{Model4a}
\begin{tabular}{|l|c|c|c|c|c|c|c|c|} \hline
	Set&$D_{4}$&	$\alpha$&	$\beta$&	$\mu$&	$\phi$&	$n_{s}$&	$r$ \\ \hline 
	&-1&	10& 	1&	1&	10.15 to&&\\
	I&&&&& 10.27&	0.968$\pm$0.004& 0.101$\pm$0.016 \\  \hline 
	&1&	1.91&&&&&\\
	II&& to 1.97& 	0.5&	1&	5.15&	0.968$\pm$0.002& 0.091$\pm$0.006 \\  \hline
		&1.01&&&&&& \\
	III&to 1.13&	2& 	0.5&	1&	5.15&	0.968$\pm$0.003& 0.092$\pm$0.008 \\	 \hline  
	&1&	1.97& 	0.497&&&&\\
	IV&&& to 0.5&	1.126&	5.15&	0.968$\pm$0.001& 0.075$\pm$0.004 \\  \hline 
	&1&	1.97& 	0.5&	1.117&&&\\
	V&&&& to 1.126&	5.15&	0.965$\pm$0.003& 0.082$\pm$0.008 \\  \hline
\end{tabular}
\end{table}

  \begin{table}[h!]
\centering
\caption{Model 4: $R+\alpha\,\ln  \left( {\frac {R}{{\mu}^{2}}} \right) +\beta\,{R}^{2}$, root2}
\label{Model4b}
\begin{tabular}{|l|c|c|c|c|c|c|c|c|} \hline
	Set&$D_{5}$&	$\alpha$&	$\beta$&	$\mu$&	$\phi$&	$n_{s}$&	$r$ \\ \hline 
	&-10.01&	1& 	1&	1&	10.06 to&&\\
	I&&&&& 10.16&	0.968$\pm$0.005& 0.081$\pm$0.014 \\  \hline 
	&10.01&	0.95&&&&&\\
	II&& to 1.01& 	1&	1&	10.15&	0.968$\pm$0.006& 0.083$\pm$0.016 \\  \hline
		&10&&&&&& \\
	III&to 10.14&	1& 	1&	1&	10.15&	0.968$\pm$0.005& 0.083$\pm$0.010 \\	 \hline  
	&10.5&	1.5& 	0.022&&&&\\
	IV&&& to 0.027&	16&	10.15&	0.968$\pm$0.009& 0.086$\pm$0.025 \\  \hline 
	&10.5&	1.5& 	0.027&	0.922&&&\\
	V&&&& to 0.931&	10.15&	0.968$\pm$0.005& 0.082$\pm$0.013 \\  \hline
\end{tabular}
\end{table}

\subsection{Model 5: $R+\alpha\, \left( {e}^{-\beta\,R}-1 \right)$}
This model is one of the simplest models that generalize the GR from two quick facts. The first one is when $\alpha$ is zero. The second one is when $\beta$ is zero so that the term in the brackets goes to zero immediately. However, there is another way to make this model reduce to GR. It can go like this: it is to apply the series expansion to the exponential term and make the free term (a term without $R$) influence as a cosmological constant.
The model is considered in \cite{nojiri2011unified,cognola2008class} as part of viable classes of $f(R)$ models in the context of describing inflation. The first derivative with respect to Ricci scalar is given as
\begin{equation}
f'=1-\alpha\,{e}^{-\beta\,R}\beta
\end{equation}
so that the scalar field becames
\begin{equation}
\phi=1-\alpha\beta\,{e}^{-\beta\,R}.
\end{equation}
The Ricci scalar is
\begin{equation}
R(\phi)=-{\frac {1}{\beta}\ln  \left( {\frac {\phi-1}{\alpha\,\beta}}
 \right) }
\end{equation}
We have $f(\phi)$ given as
\begin{equation}
f(\phi)=-{\frac {1}{\beta}\ln  \left( {\frac {\phi-1}{\alpha\,\beta}}
 \right) }+\alpha\,\left( {\frac {\phi-1}{\alpha\,\beta}} -1 \right) \; .
\end{equation}
With the potential $V(\phi)$ being
\begin{equation}
V(\phi)={\frac {1}{12\beta} \Big[  \left( 2\,{\phi}^{2}-8\,\phi+6
 \right) \ln  \left( {\frac {-\phi+1}{\alpha\,\beta}} \right) -(8\alpha\,\beta-14)\phi
-5\,{\phi}^{2}-5 \Big] }
+D_{6}\; ,
\end{equation}
where $D_{6}$ is a constant of integration.

  \begin{table}[]
\centering
\caption{Model 5: $R+\alpha\, \left( {e}^{-\beta\,R}-1 \right)$}
\label{Model5}
\begin{tabular}{|l|c|c|c|c|c|c|c|} \hline
	Set&$D_{6}$&	$\alpha$&	$\beta$&	$\phi$&	$n_{s}$&	$r$ \\ \hline 
	&0.3&	25& 	1&	0.22 to&&\\
	I&&&& 0.33&	0.969$\pm$0.005& 0.083$\pm$0.015 \\  \hline 
	&0&	13&&&&\\
	II&& to 25& 	1&	0.2&	0.968$\pm$0.005& 0.085$\pm$0.015 \\  \hline
		&0.3&&&&& \\
	III&to 1.2&	15& 1.25&	0.3&	0.968$\pm$0.006& 0.086$\pm$0.016 \\	 \hline  
	&0.5&	15& 	0.65&&&\\
	IV&&& to 1.2&	0.2&	0.967$\pm$0.007& 0.087$\pm$0.019 \\  \hline 
\end{tabular}
\end{table}

\subsection{Model 6: $R-\left( 1-n \right) {\mu}^{2} \left( {\frac {R}{{\mu}^{2}}} \right)^{n}$ }
This model also is one of the generalization of GR.
The quick aspect of this model that one can easily notice is that when $n=1$, this model directly reduces to GR.
This model was considered in \cite{scalar3,Carroll1}, where the study of the effect of Chameleon mechanism is done. In this $f(R)$ model, one has
\begin{equation}
f'=1-{\frac { \left( 1-n \right) {\mu}^{2}n}{R} \left( {\frac {R}{{\mu}^{2}}} \right) ^{n}},
\end{equation}
So that the use of definition of the scalar field yields
\begin{equation}
\phi=1-{\frac { \left( 1-n \right) {\mu}^{2}n}{R} \left( {\frac {R}{{\mu}^{2}}} \right) ^{n}}\; .
\end{equation}
The Ricci scalar takes the form
\begin{equation}
R(\phi)={\mu}^{2}\left( {\frac {\phi-1}{n \left(-1+n \right) }} \right)^{\frac {1}{-1+n}}  .
\end{equation}
Then $f(\phi)$ is given as
\begin{equation}
f(\phi)={\mu}^{2}\left( {\frac {\phi-1}{n \left(-1+n \right) }} \right)^{\frac {1}{-1+n}} 
- \left( 1-n \right) {\mu}^{2}\left( {\frac {\phi-1}{n \left(-1+n \right) }} \right)^{\frac {n}{n-1}} \;.
\end{equation}
The potential  $V(\phi)$ is given as
\begin{equation}
\begin{split}
&V(\phi)=\Big[\frac {2(-n^{2}+2n-2)(1-\phi)\mu^{2}}{3(2\,n-1)}-\frac {2{\mu}^{2}n(1-n)}{3 \left( -1+n \right) ^{-1}+1} \Big]\left(\frac {\phi-1}{n \left(n-1 \right) }\right)^{\frac{n}{n-1}}\\
 &+\Big[\frac{(n-2+\phi+\phi^{2}-n\phi^{2})\mu^{2}}{3(2n-1)}+\frac{(1-\phi)\mu^{2}}{3n(2n-1)}\Big]\left(\frac {\phi-1}{n \left(n-1 \right) }\right)^{\frac{1}{n-1}}+D_{7}\;,
\end{split}
\end{equation}
where $D_{7}$ is a constant of integration.
\begin{table}[]
\centering
 \caption{Model 6: $f(R)=\rm{R-(1-n)\mu^2 \bigg(\frac{R}{\mu^2}\bigg)^n}$}
\label{Model6}
\begin{tabular}{|l|c|c|c|c|c|c|c|c|} \hline
Set&$D_{7}$&$\mu$&n&$\rm{\phi}$&	$\rm{n_s}$&	$r$  \\ \hline
&0.9&8 to&&&&\\
I&& 9.6&1.5&1.5&0.968$\pm$ 0.004&	0.243$\pm$0.016   \\ \hline
&0.9&10.6&1.5&1.376&& \\
II&&&&to 1.4&0.970$\pm$ 0.003&	0.301$\pm$0.003  \\ \hline
&1.7 to&&&&&\\
III& 2.2 &15.6&1.5&1.4&0.967$\pm$ 0.005&	0.305$\pm$0.021  \\ \hline
&1.7 &15.6&1.5 to&&&\\
IV&&& 1.73&1.4&0.969$\pm$ 0.003&	0.241$\pm$0.044 \\ \hline
\end{tabular}
\end{table}
\subsection{Model 7: $R+\,{\frac {{R}^{2}}{6{M}^{2}}}$}
This model is one of the Starobinsky models and was studied in \cite{8} and its generalization was considered in \cite{2018arXiv180803087V}. Depending on the way the free term $M$ is defined, this model can be considered as a subclass of the $f(R)$ model considered in Subsection \ref{linearMODELLLL} with appropriate definition of $\alpha$ and $\beta$ in that model. The first derivative of this Lagrangian with respect to $R$ gives
\begin{equation}
f'=1+{\frac {R}{3{M}^{2}}}\;,
\end{equation}
so that the scalar field be
\begin{equation}
\phi=1+{\frac {R}{3{M}^{2}}}\;.
\end{equation}
We get the Ricci scalar as a function of the scalar field as
\begin{equation}
R(\phi)=3\, \left( \phi-1 \right) {M}^{2} \;.
\end{equation}
Then $f(\phi)$ is given as
\begin{equation}
f(\phi)=3\, \left( \phi-1 \right) {M}^{2}+\frac{3}{2}\left( \phi-1 \right) ^{2}{M}^{2} \;.
\end{equation}
The potential  $V(\phi)$ is given as
\begin{equation}
V(\phi)=-{M}^{2} \left( \frac{1}{3}{\phi}^{3}-\frac{1}{2}{\phi}^{2} \right) +2\,{M}^{2}
 \left( \frac{1}{2}{\phi}^{2}-\phi \right) +\frac{1}{3}\left( \phi-1 \right) ^{3}
{M}^{2}+D_{8}\;,
\end{equation}
where $D_{8}$ is a constant of integration.

\begin{table}[h!]
\centering
\caption{Model 7: $f(R)=R+\frac{R^{2}}{6M^{2}}$}
\label{Model7}

\begin{tabular}{|l|c|c|c|c|c|c|c|c|c|} \hline
 Set&$D_{8}$ & M & $\phi$ &  $n_{s}$     & $r$       \\ \hline
	&5000&	100 &	0.40&& \\
	I&&&to 0.66&	0.963$\pm$ 0.004&	0.058$\pm$ 0.011 \\	 \hline	
	&5000&	91 to&&&\\
	 II&&95.5 &	0.65&	0.967$\pm$ 0.004&	0.0409$\pm$ 0.008  \\	 \hline
	&300 to &&&&\\
III&	1200&95.5 &	1.85&	0.968$\pm$ 0.004&	0.060$\pm$ 0.009  \\	 \hline
\end{tabular}
\end{table}

\section{Discussions}\label{DISCUSSIONSSS}

\noindent The two inflationary parameters are computed in this work, namely scalar spectral index $n_{s}$ and tensor-to-scalar ratio $r$. These parameters can be obtained observationally \cite{Spergel1,Spergel2}.
Here we summarize the results of the two parameters $n_{s}$ and $r$ from six survey reports, as shown in Table \ref{Surveyys}.

\begin{table}[h!]
\centering
\caption{Values of $n_{s}$ and $r$ from different surveys}
\label{Surveyys}
\begin{tabular}{|l|c|c|c|c|c|c|c|c|c|} \hline
 &I \cite{dunkleyfive}& II \cite{larson2011seven}&III \cite{hinshaw2013nine}&IV \cite{tegmark2004cosmological}&V \cite{ade2015joint}&VI \cite{ade2016planck}     \\ \hline
 $n_{s}$&0.963$^{+0.014}_{-0.015}$& 0.972$^{+0.014}_{-0.014}$ & 0.972$^{+0.013}_{-0.013}$ &0.97$^{+0.12}_{-0.10}$ &0.968$^{+0.006}_{-0.006}$&0.968$^{+0.006}_{-0.006}$  \\ \hline
 $r$&$<$0.43	 &$<$0.36 & $<$0.13&$<$0.5 &$<$0.12 &$<$0.11  \\ \hline
\end{tabular}
\end{table}
\noindent For example in the full Planck Survey results reported in \cite{ade2016planck}, it is shown that the spectral index $n_{s}$ is $0.968\pm 0.006$. The same value is reported in \cite{ade2015joint}. The upper limit for the tensor-to-scalar ratio $r$ is $<0.11$. But in \cite{ade2015joint}, it was reported that  $r<0.12$. However, in \cite{hinshaw2013nine}, the Nine-year WMAP (Wilkson Microwave Anisotropy Probe) reported $r<0.13$. The remaining surveys reported a number far bigger than these.  
In this work, we numerically computed those parameters and the obtained values are compared with the recent values from the  surveys \cite{dunkleyfive,larson2011seven,hinshaw2013nine,
tegmark2004cosmological,ade2015joint,ade2016planck}.
If one carefully looks in Table \ref{Surveyys}, one could find that the minimum averaged value of $n_{s}$ is reported in \cite{dunkleyfive}, a five-year WMAP report. The maximum averaged number for $n_{s}$ is reported in \cite{hinshaw2013nine}. We therefore considered any model that produce $n_{s}$ value less than the above minimum  range or beyond the maximum range to be not supported by observations.
  
\noindent Note that we have selected different ranges for free parameters that are available in each toy model to form a set. 
For example, in the $\beta R^{n}$ toy model, the available free parameters are $\beta,n$. We did the same for all other models and we report the results as follows.
The results for the power-law model, $\beta R^{n}$ for different ranges of the free parameters are in the range predicted by the observations but for $r$, some of the values are out of the predicted number. However, the good numbers are the ones in Set $V$, which produces $n_{s}=0.9683$ and $r=0.035$ as average values.  
For the $f(R)=\alpha R+\beta R^{n}$ model, we have results of $r$ values that are in the range predicted by observations for both changing $\phi$ and changing $\alpha$. The values for $n_{s}$ for all possible parameter space considered are in range of the ones from the observations. Though values of $r$ are a bit bigger, Set $I$ has good values ($n_{s}=0.969$ and $r=0.175$) in comparison with other sets.
The results for $\alpha\,{e}^{\lambda\,R}$ model are also in ranges suggested by the observations if one considers the values of $n_{s}$ and some of the $r$ values. In this model, the good set is Set $IV$ since it has the smallest value of $r=0.149$ and corresponding $n_{s}=0.969$. Set $III$ has $n_{s}$ value which is below the minimum value reported by the observations.
For the $R+\alpha \ln(R/\mu^2)+\beta R^2$, we have two different roots for $R(\phi)$ solution. This results in two different type results. The results that are in connection with the first root are presented in Table \ref{Model4a}. The results of the second root are presented in Table \ref{Model4b}. However, for the results in Table \ref{Model4a}, the values that are located in the Set $V$ produce averaged value of $n_{s}$ which is below the averaged value reported in the Planck Survey results. The remaining sets have averaged values of $n_{s}$ that match with the observed value. The averaged values of $r$ for all sets are in good range in comparison to the observed values. For the second root (see Table \ref{Model4b}), all sets have averaged values of both $n_{s}$ and $r$ which are in range suggested by the observations.
For the model the form $R+\alpha (e^{-\beta R}-1)$, we have results for $n_{s}$ that are in good agreement with the observations and two sets have the same value as the Planck survey results. The averaged values of $r$ is in range predicted by the recent observations.
The results for $R-\left( 1-n \right) {\mu}^{2} \left( {\frac {R}{{\mu}^{2}}} \right)^{n}$ model are in good range but all the values of $r$ are above $0.13$. However, one can choose Set $IV$ as being the closest set in comparison to other sets with $n_{s}$ value of $0.969$ and minimum $r$ with value of $0.197$. 
Some of the results for $R+1/6\,{\frac {{R}^{2}}{{M}^{2}}}$ model for varying any free parameter, are in ranges as the observations suggest, but Set $I$ has $n_{s}$ which below the minimum value reported from the observation, see Table \ref{Model7}. This model has the best set as Set $III$ with averaged values $n_{s}=0.968$ and $r=0.06$.
 
\section{Conclusions}\label{CONCLU}
In this work, we have considered the relationship between two theories namely, $f(R)$ and scalar tensor theory of gravity. We have considered several $f(R)$ toy models with interest in the inflationary universe. We have numerically computed two inflationary parameters namely scalar spectral index $n_{s}$ and tensor-to-scalar ratio $r$. For each $f(R)$ toy model, we have computed the error to the averaged value and the upper and lower bounds are used to compare the results with observational survey with the most recent being the Planck Survey results reported in \cite{ade2016planck}. For each model, we have explored different ranges of the free parameters as classified in Sets. For each model, we have selected a Set with averaged values that are close to the recent Planck Survey report. In general, all considered toy models produced good averaged values of $n_{s}$. But some of them produced values of $r$ that are above the predicted result of the 2015 Planck survey. For instance, the $R-\left( 1-n \right) {\mu}^{2} \left( {\frac {R}{{\mu}^{2}}} \right)^{n}$ model has the $r$ values larger than $0.13$, a value presented by the nine-years WMAP survey. We inspected that the model $R+\alpha \ln(R/\mu^2)+\beta R^2$,
see Tables \ref{Model4a} and \ref{Model4b}, produced, in all considered Sets, values of both $n_{s}$ and $r$ that are in range suggested by the observations.
However the extended free parameter space, for a particular $f(R)$ model, is needed to be covered to generally decide on the constraints of $f(R)$ models based on the current and coming observations.
The current exercise aims to only demonstrate that one can in principle constrain a given $f(R)$ toy model from the consideration of cosmological inflation. This is made possible due to the definition of the inflation field as given in Eq.\eqref{scalarfield}.

\section*{Acknowledgement}
JN and JM gratefully acknowledge financial support from the Swedish 
International Development Cooperation Agency (SIDA) through the International Science Program
(ISP) to the University of Rwanda (Rwanda Astrophysics, Space and Climate Science Research Group), project number RWA01. SS acknowledges the financial support from Wolkite University. JN, JM and SS are grateful to the Cosmology Lectures offered by Prof David F. Mota in the ISYA-schools (2017/2018).
AA acknowledges that this work is based
on the research supported in part by the National Research Foundation (NRF) of
South Africa.
\bibliographystyle{unsrt}
 
\end{document}